\documentstyle[12pt,aaspp]{article}
\begin{document}
\title{
LARGE SCALE STRUCTURE EFFECTS ON
THE GRAVITATIONAL LENS  
IMAGE POSITIONS AND TIME DELAY} 
\slugcomment{{\it Astrophysical Journal Letters}, in press}

\author{Uro\v s Seljak\altaffilmark 1}
\affil{Department of Physics, MIT, Cambridge, MA 02139 USA}
\begin{quote}
\altaffilmark{}
\altaffiltext{1}{Also Department of Physics, University of Ljubljana, 
Jadranska 19, 61000 Ljubljana, Slovenia}
\end{quote}

\def\bi#1{\hbox{\boldmath{$#1$}}}
\begin{abstract}
We compute the fluctuations in gravitational lens image positions and 
time delay caused by large scale structure correlations. We show that 
these fluctuations can be expressed as a simple integral over the density
power spectrum.  Using the {\sl COBE} normalization we find that positions 
of objects at cosmological distances are expected to deviate from their 
true positions by a few arcminutes.  These deflections are not directly 
observable. The positions of the images relative to one another fluctuate 
by a few percent of the relative separation, implying that one does not 
expect multiple images to be produced by large scale structures. Nevertheless, 
the fluctuations are larger than the observational errors on the positions 
and affect reconstructions of the lens potential.  The time delay fluctuations 
have a geometrical and a gravitational contribution. Both are much larger than 
the expected time delay from the primary lens, but partially cancel each other.
We find that large scale structure weakly affects the time delay and time delay
measurements can be used as a probe of the distance scale in the universe.
\end{abstract}

\section{Introduction}
The possibility of studying the physical parameters of the distant universe
using gravitational lenses (GL) was first
suggested in the 1960s.
In particular, Refsdal (1964, 1966) suggested that one could 
determine the masses of galaxies and the 
Hubble constant using the observed image properties, most notably
their positions, magnifications and time delays
between images of the same source. 
%Key assumption underlying the method
%is that the observed properties are not significantly 
The latter became especially interesting after the  
time delay in the system 0957+561 was measured (e.g. Vanderriest et al.
1989; Leh\' ar et al. 1992)
and a value of $H_0$ derived (Rhee 1991;
Roberts et al. 1991).
Alcock \& Anderson (1985, 1986), 
Watanabe, Sasaki \& Tomita (1992) and Sasaki (1993)
criticized the method, arguing that 
large scale structure (LSS) might significantly affect the 
time delay. Unfortunately, 
their arguments were only qualitative and could not give
realistic predictions of the amplitude of fluctuations. 
For example, Falco, Gorenstein \& Shapiro (1991) found that the
effect Alcock \& Anderson discussed 
can change the derived value of
$H_0$ by only a few percent. 
In light of this 
many workers in the field have taken 
an optimistic view and assumed that the derived value of $H_0$
gives at least an upper limit to the actual value (e.g. 
Borgeest \& Refsdal 1984). 
These arguments are based on the fact that 
mass density is always positive and therefore always focuses the 
rays. However, this is only correct in Newtonian gravity 
and becomes invalid in cosmological applications, where underdensities
such as voids give an effective negative mass density (e.g. Nityananda \&
Ostriker 1984). In general, the question whether 
LSS could significantly affect the measured 
properties of the lens has remained largely unanswered. 

In this paper we present a  
calculation of position and time delay dispersions 
in a GL system. The calculation
is done using linearized general relativity for 
some realistic {\sl COBE} normalized cosmological models.
In \S 2 we present the method, based on the geodesic equation
for a perturbed Robertson-Walker metric. 
A similar approach has been used by
Linder (1990) and
Cay\' on, Mart\' inez-Gonz\' alez \& Sanz (1993a,
1993b) 
to study the GL effects on the cosmic microwave background 
and by Kaiser (1992) 
to derive the ellipticity 
correlation function of distant galaxies.
An alternative method, based on optical scalars, has been developed
by Gunn (1967) and applied to the ellipticity and magnification
correlation function calculations by Babul \& Lee (1991), 
Blandford et al. (1991) and 
Miralda-Escud\' e (1991). 
In \S 3 we apply the method to compute the 
fluctuations in
the image position relative to the
true position and relative to another image position.
In \S 4 we present the calculation of fluctuations in time delay.
In \S 5 we present the conclusions and comment on their agreement
with previous work on this subject.

\section{Formalism}
Our framework is a perturbed flat Robertson-Walker
model with small-amplitude scalar metric fluctuations. In the 
longitudinal gauge
(Bardeen 1980; Mukhanov, Feldman and Brandenberger
1992) one can write the line element as 
%\footnote[1]{We use c=1}
\begin{equation}
ds^2=a^2(\tau)\biggl[-(1+2\phi)d\tau^2+(1-2\phi)
d\bi{x}\cdot d\bi{x}\biggl]\ .
\label{metric}
\end{equation}
We assume that $\vert \phi \vert \ll 1$ and neglect all the terms of order 
$O(\phi^2$) and higher. This is a good approximation almost everywhere in the 
universe,
except near black holes.
We adopt units such that $c=1$.
In the line element above we neglected the contributions from 
vector and tensor modes and the gravitational 
effects from anisotropic stresses. These approximations are valid
nearly always, especially in the regime of interest for us, which is 
the matter dominated era with
fluctuation wavelengths small compared to the 
Hubble distance. 
In this case $\phi$ can be interpreted as the Newtonian
potential and, neglecting the contributions from 
wavelengths larger than the Hubble distance,
it obeys the cosmological Poisson equation
\begin{equation}
\nabla^2\phi=4\pi Ga^2(\rho-\bar{\rho}) , 
\label{poisson}
\end{equation}
where $\rho$ is the local density and $\bar{\rho}$ the mean
density in the universe (e.g. Bertschinger 1993).
We denote the time dependence of the potential 
with $F(\tau)$, which is independent 
of the scale in linear perturbation theory, assuming the dominant 
matter component has negligible Jeans length.
For the zero curvature
model with 
a cosmological constant, $\Omega_m+\Omega_{\lambda}=1$, one has
(Heath 1977) 
\begin{equation}
F(a)={\sqrt{\Omega_m+\Omega_{\lambda}a^3} \over a^{5/2} }
{\int_0^{a}X^{3/2}da \over \int_0^1 X^{3/2}da },
\end{equation}
where $X=a/(\Omega_m+\Omega_{\lambda}a^3)$ and $H_0\tau=\int_0^a da/
(\Omega_ma+\Omega_{\lambda}a^4)^{1/2}$; we normalize $a(\tau)=1/(1+z)$
to $a=1$ today. $H_0$ is the Hubble 
constant and $\Omega_m$ and $\Omega_{\lambda}$
are the matter and vacuum densities, respectively, 
in units of the critical density.
The simplest model has $\Omega_m=1$, for which $F(\tau)=1$
and $a(\tau)=(H_0\tau/2)^2$. 

\begin{figure}[p]
\vspace*{5 cm}
\caption{Photon propagation relative to the source-observer line.} 
\includegraphics{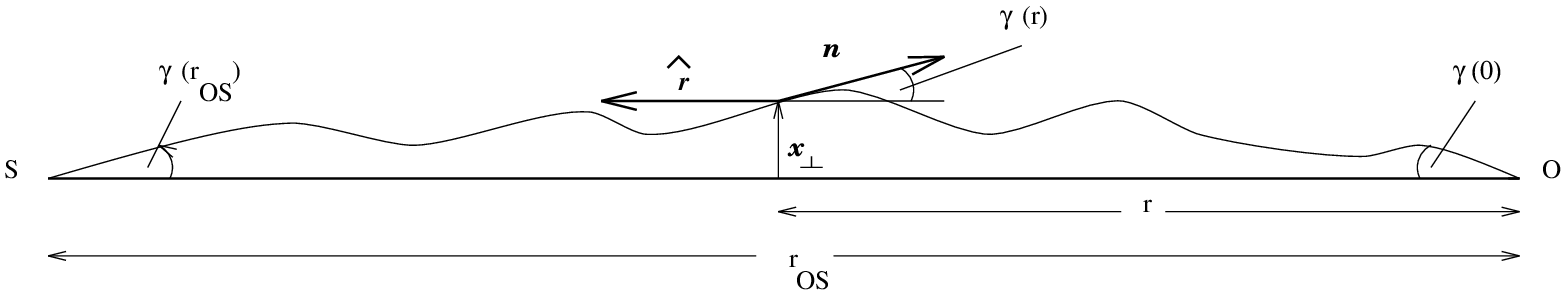}

\label{fig0}
\end{figure}

Suppose a photon is emitted from a distant source
toward the observer (see Figure \ref{fig0}).
If there are no perturbations present the photon will travel along a
null geodesic in the radial direction
\footnote[2]{Here and throughout the paper 
all 3-vectors are defined in the
unperturbed
comoving coordinates of 3-space, which is a hypersurface of constant 
$\tau$. The geometry of this 3-space is a simple Euclidean geometry.} 
, $\bi{x}(r)=r\bi{\widehat{r}}$ 
and $r(\tau)=\tau_0-\tau$, where $\tau_0$ is the conformal time today.
Adding a perturbation changes the photon trajectory into 
$\bi{x}(r)=r\bi{\widehat{r}}+\bi{ x_{\perp}}(r)$, where $\bi{ x_{\perp}}(r)$
is the excursion of the photon in the direction orthogonal to
$\bi{\widehat{r}}$. 
The 3-tangent to the 
geodesic curve is given by $\bi{n}=d\bi{x}/dl$, where $dl=(d\bi{x}\cdot
d\bi{x})^{1/2}$ is the comoving path length along the geodesic. 
Null geodesics obey the relation $d\tau=(1-2\phi)dl$. Even when 
metric perturbations are present, we can 
continue to parametrize the geodesic with the unperturbed radial 
coordinate $r$. The relation between $r$ and $l$ is given by
$dr/dl=\bi{n}\cdot \bi{\widehat{r}}$.
The linearized 
space part of the photon geodesic equation derived from the metric in equation 
(\ref{metric})
gives the rate of change in photon direction, 
\begin{equation}
{d\bi{n} \over dl}=2\bi{n} \times (\bi{n} \times {\bf \nabla}
\phi)
\label{dndl}
\end{equation}
(e.g. Weinberg 1972 eq. 9.2.6-7; note that there is a factor of 2 
missing in eq. 9.2.7). For weak gravitational fields ($\vert \phi \vert \ll
1$),  
$\phi$ can be viewed as providing a force deflecting the photons
and affecting their 
travel time while they 
propagate through unperturbed space-time. 
We shall adopt this quasi-Newtonian interpretation throughout this paper.

Because $\bi{n}$ is normalized, it is sufficient to consider only its two 
components orthogonal to $\bi{\widehat{r}}$. We define the two-dimensional 
photon direction angle 
$\bi{\gamma}(r)=[\gamma_1(r),\gamma_2(r)]$
relative to the unperturbed photon 
direction $-\bi{\widehat{r}}$ (taken as the third direction)
as $\bi{n}=(\gamma_1, \gamma_2,-1)$.
We assume small deflection angles (this
will be justified in \S 3). 

The evolution of $\bi{\gamma}(r)$ is given by
\begin{equation}
\bi{\gamma}(r)=\bi{\gamma}(r_{OS})+\bi{\alpha_{\perp}}(r,r_{OS}),
\end{equation}
where $\bi{\gamma}(r_{OS})$ is the initial photon direction 
(here $r_{OS}$ is
the comoving distance between the observer and the source)
and
$\bi{\alpha_{\perp}}(r,r_{OS})=[\alpha_{\perp,1}(r,r_{OS}),
\alpha_{\perp,2}(r,r_{OS})]$ is a two-dimensional deflection angle
produced by the potential along the geodesic between the source 
and the point under consideration at $r$. 
From equation (\ref{dndl}) follows
\begin{equation}
\bi{\alpha_{\perp}}(r,r_{OS})=\int_{r}^{r_{OS}} 
\bi{g_{\perp}}(r)dr. 
\label{gamdef}
\end{equation}
We introduced $\bi{g_{\perp}}(r)$ defined as
\begin{equation}
\bi{g_{\perp}}(r)=-2{\bf \nabla_{\perp}}
\phi[\bi{x}(r),\tau(r)]=-2[{\bf \nabla}-\bi{\widehat{r}}({\bf \nabla}
\cdot \bi{\widehat{r}})]\phi[\bi{x}(r),\tau(r)].
\end{equation}
The photon excursion in the plane perpendicular to 
$\bi{\widehat{r}}$ 
is given by
\begin{equation}
\bi{x_{\perp}}(r)=\int_r^{r_{OS}}\bi{\gamma}(r)dr.
\end{equation}
The initial photon direction
$\bi{\gamma}(r_{OS})$ must be chosen so that 
$\bi{x_{\perp}}(0)=0$, 
i.e., the ray must pass through the observer's position. 
This gives the lens equation
$$
\int_0^{r_{OS}} \bi{\gamma}(r)dr=0 
$$
or
\begin{equation}
r_{OS}\bi{\gamma}(r_{OS})=
-\int_0^{r_{OS}}\bi{\alpha_{\perp}}(r,r_{OS})dr.
\label{lenseq} 
\end{equation}
This lens equation is valid for an arbitrary mass distribution between 
the source and the observer.
It cannot be solved explicitly in general, because 
$\bi{\alpha_{\perp}}(r,r_{OS})$ 
depends on the unknown initial photon direction $\bi{\gamma}(r_{OS})$.
Instead, one has to solve an integral equation
using, for example, the ray-shooting method.

We may also ask what is the photon direction at
the observer's position so that $\bi{x_{\perp}}(r_{OS})=0$.
This is given by the
image position angles $\bi{\gamma}(0)$. 
For the special case of a single thin lens one has
\begin{equation}
\bi{\alpha_{\perp}}(r,r_{OS})= \left\{ \begin{array}{ll}
0, & r>r_{OL} \\
\bi{\widehat{\alpha}}, & r\leq r_{OL},
\end{array}
\right.
\end{equation}
where $\bi{\widehat{\alpha}}$ is the bending angle in 
the lens plane and
$r_{OL}$
is the comoving 
distance between the observer and the lens.
Equation (\ref{lenseq}) then gives 
\begin{equation}
r_{OS}\bi{\gamma}(0)=-r_{LS}\bi{\widehat{\alpha}},
\label{lenseq2}
\end{equation}
which is the usual
lens equation in the thin lens approximation 
(e.g. Schneider, Ehlers \& Falco 1992; Blandford \& Kochanek 1987).
Here $r_{LS}$ is the comoving distance between 
the observer and the source and
$r_{OS}=r_{OL}+r_{LS}$. 

The distances can be calculated 
for a given cosmological model
from the
source redshift $z_s$ and lens redshift $z_l$ (assuming
that the two redshifts can be measured). 
For example, for $\Omega_m=1$ one has $r_{OL}=2H_0^{-1}[1-(1+z_l)^{-1/2}]$ and
$r_{LS}=2H_0^{-1}[(1+z_l)^{-1/2}-(1+z_s)^{-1/2}]$. We use the
unperturbed comoving distance-redshift relation, because the deviations 
from it are $O(\phi)$ and can be neglected to the first 
order.  
Note that we expressed the lens equation
using comoving distances. One can reexpress it, if one wishes, with 
angular diameter distances using the relation  
$d_{AS}=r_{AS}/(1+z_S)$, where $z_S$ is the redshift of the 
source and 
$d_{AS}$ and $r_{AS}$ are, respectively, the angular diameter distance 
and comoving distance between the point A and the source.

Similarly, the net time delay along the photon path relative to the 
unperturbed path is given by 
\begin{eqnarray}
\Delta t=\Delta \tau(a=1)=\int_0^{r_{OS}}\biggl[{1-2\phi(r)\over\bi{n \cdot 
\widehat{r}}}-1 \biggl]dr \approx \nonumber \\ 
\int_0^{r_{OS}}\biggl[ {1 \over 2}\bi{\gamma}^2(r)-
2\phi[\bi{x}(r),\tau(r)]\biggl]dr,
\label{tdgen}
\end{eqnarray}
where $\bi{\gamma}^2(r)=\gamma_1^2(r)+\gamma_2^2(r)$ and we used the small 
angle approximation; the unperturbed null geodesic equation may be used
for $\tau(r)=\tau_0-r$.
The first and second 
term in equation (\ref{tdgen}) are the geometrical and gravitational 
(also called potential or Shapiro)
time delay contribution, respectively.
Again, 
for the particular case of the thin lens approximation this reproduces 
the usual time delay equation, as shown in \S 4.

Although we have derived our equations using the geodesic equations of a 
perturbed Robertson-Walker metric, the final expressions agree with the usual 
thin lens equations, when we interpret $\phi$ as the Newtonian potential, 
obeying equation \ref{poisson}. 
In addition, one
does not have to assume
that density perturbations are small,
as long as $\vert \phi \vert \ll 1$. 
These are advantages of the longitudinal 
gauge compared to other gauge choices (e.g. synchronous gauge).
The particular strength of the approach presented here
is that it remains valid in cosmological applications and 
can give the deflection
angle and time delay due to any matter distribution between the source and
observer. 
Use of comoving coordinates and conformal time  
greatly simplifies the GL
equations. In particular, 
all the equations above use a simple Euclidean spatial geometry and
there is no need to use angular
distances or worry about expansion factors.
The general expressions above would be more cumbersome when 
expressed with angular diameter distances.

We are interested in LSS effects on the image
properties of lenses. 
Because of the statistical nature of cosmological
theories one can only predict the ensemble averages of a given 
quantity, such as its mean and variance. All the GL
effects are given through the 
gravitational potential $\phi[\bi{x},\tau(r)]$. Its statistical 
properties can best be described with the Fourier transformed
potential $\phi(\bi{k})$, 
\begin{equation}
\phi[\bi{x},\tau(r)]=\int d^3k \phi(\bi{k})e^{i\bi{k\cdot x}}F[\tau(r)].
\label{ft}
\end{equation}
The ensemble mean and variance of the Fourier transform of the potential 
are $\langle \phi(\bi{k})\rangle=0$
and $\langle \phi(\bi{k})\phi^*(\bi{k'})\rangle=P_{\phi}(k)\delta^3(\bi{
k-k'})$, where $P_{\phi}(k)$ is the power spectrum of the potential
(see e.g. Bertschinger 1992). We will use the power spectrum of the 
potential in this paper because it leads to the simplest expressions, but 
note that one can easily reexpress the results
with the density power spectrum 
using equation \ref{poisson}.

A particularly convenient
normalization of $P_{\phi}(k)$ is given by the cosmic microwave 
background anisotropy measured by 
{\sl COBE} - the quadrupole $Q_2=(\Delta T/T)_2=(6\pm 1) \times 10^{-6}$
(Smoot et al. 1992; Seljak \& Bertschinger 1993).
On the large scales
probed by {\sl COBE} the dominant contribution to $\Delta T/T$ is given
by the Sachs-Wolfe (1967) effect, induced by the same metric fluctuations
that cause the fluctuations in time delay and image positions.
Assuming no tensor mode contribution to CMB anisotropies
and isentropic (adiabatic) fluctuations
we can express the quadrupole in terms of the
power spectrum of the potential (Bond \& Efsthathiou 1987) as
\begin{equation}
Q_2^2={20\pi K_2^2 \over 9} \int_0^{\infty}k^2P_{\phi}(k)j_2^2(2k/H_0)dk.
\end{equation}
Here $j_2(x)$ is the spherical Bessel function of order 2 and
$K_2^2 \geq 1$ is the amplification coefficient
due to the time dependent
potential (Kofman \& Starobinsky 1985).
If $\Omega_m=1$ the potential is time independent and $K_2^2=1$.
For the scale-invariant
Peebles-Harrison-Zel'dovich spectrum
$P_{\phi}(k)= Ak^{-3}$ one gets
\begin{equation}
Q_2^2={5\pi K_2^2A\over 27}.
\label{q2}
\end{equation}

\section{Fluctuations in Angular Position}

The first question we will ask is what is the rms fluctuation in the 
photon direction at the observer's position relative to the unperturbed
direction. 
This is defined as
\begin{equation}
\sigma_{\gamma}={1 \over \sqrt{2}}\langle
\bi{\alpha_{\perp}}(0,r_{OS})\cdot \bi{\alpha_{\perp}}(0,r_{OS})
\rangle^{1/2},
\label{varalpha}
\end{equation}
where $\bi{\alpha_{\perp}}(0,r_{OS})$ is the total deflection angle accumulated 
between the source and the observer, as defined in equation (\ref{gamdef}).
In order to compute the variance we will make an additional 
assumption that 
the statistical properties of the potential sampled  
by the perturbed photon geodesic are approximately equal to those 
along the unperturbed one, 
\begin{equation}
\langle
\phi[\bi{x},\tau(r)]
\phi[\bi{x},\tau(r)]
\rangle
\approx 
\langle
\phi[r\bi{\widehat{r}},\tau(r)]
\phi[r\bi{\widehat{r}},\tau(r)]
\rangle.
\label{assum}
\end{equation}
This assumption will be discussed later.

We want to compute statistical properties 
of the transverse gradient
of the potential field  $\phi$ 
integrated along line of sight in the direction $\bi{\gamma}$ 
and weighted with a function $q(r)$, 
\begin{equation}
\bi{p}(\bi{\gamma})=-2\int_0^{r_0}[\bi{\nabla_{\perp}}\phi(r,\bi{\gamma})]
q(r)dr. 
\end{equation}
We will use the Fourier space analog of Limber's
equation (e.g. Kaiser 1992) to calculate the correlation function 
from the power spectrum, 
\begin{equation}
C(\gamma= \vert \bi{\gamma}_1-\bi{\gamma}_2 \vert
)=\langle \bi{p}(\bi{\gamma}_1)\cdot \bi{p}(\bi{\gamma}_2)\rangle
=8\pi^2\int_0^{\infty}k^3dk\int_0^{r_0}P_{\phi}[k,
\tau(r)]q^2(r) J_0(kr\gamma)dr ,
\label{limber}
\end{equation}
where $J_0(x)$ is the Bessel function of order 0.
Equation \ref{limber} assumes that the scales contributing to 
the correlation function are much smaller than the photon 
pathlength $r_0$. As shown below, this is usually 
satisfied in most models of LSS
for sources at cosmological distances. No assumption on the 
power spectrum has been made and equation \ref{limber}
can be used both in linear and in non-linear regime.
To compute the variance  
$\sigma_{\gamma}$  
we use the above expression with $\gamma=0$,
$q(r)=1$ and
$r_0=r_{OS}$. 
Assuming linear evolution and $\Omega_m=1$ we obtain
\begin{equation}
\sigma_{\gamma}=
\left( 8\pi^2r_{OS}\int_0^{\infty}P_{\phi}(k)k^3dk\right)^{1/2}.
\label{varalphafinal}
\end{equation}

The rms fluctuation in photon 
direction at the distance $r_{OS}$ from the source,  $\sigma_{\gamma}$, 
is related to the
rms angular 
fluctuation of the true 
source position relative to the observed position. 
To compute this we have to back-propagate all the 
photons with a fixed final direction and ask what are their angular
excursions in the source plane, 
\begin{equation}
\sigma_{\theta}={1 \over \sqrt{2}}\biggl\langle\biggl[
{1 \over r_{OS}}\int_0^{r_{OS}}\bi{\alpha_{\perp}}(0,r)dr \biggl]\cdot
\biggl[ {1\over r_{OS}}\int_0^{r_{OS}}\bi{\alpha_{\perp}}(0,r)dr \biggl]
\biggl\rangle^{1/2}.
\label{sth}
\end{equation}
Integrating by parts the two terms in equation (\ref{sth}) we find that
$\sigma_{\theta}$ is given by a similar expression 
as $\sigma_{\gamma}$ with $q(r)=(1-r/r_{OS})$. Using equation
\ref{varalphafinal} gives 
\begin{equation}
\sigma_{\theta}={\sigma_{\gamma} \over \sqrt{3}}=
\left({8\pi^2r_{OS} \over 3}
\int_0^{\infty}P_{\phi}(k)k^3dk\right)^{1/2}.
\label{vargammafinal}
\end{equation}

To get some intuition about the scaling of the 
amplitude with the parameters we will present
an estimate of the fluctuations
for a particularly simple power spectrum approximating
inflationary models with a physical transfer function:
\begin{equation}
P_{\phi}(k)= \left\{ \begin{array}{ll}
Ak^{-3} &, k<k_0\\
Ak^{-7}k_0^{4} &, k>k_0
\end{array}
\right.
\label{ps}
\end{equation}
The spectral indices have been chosen to agree with the cold dark matter 
model in the
limits of small and large $k$.
Applying this power spectrum to equations (\ref{varalphafinal}) and
(\ref{vargammafinal}) we
find 
\begin{equation}
\sigma_{\theta} = {\sigma_{\gamma}\over \sqrt{3}}
 \approx 7 Q_2 (k_0 r_{OS})^{1/2}. 
\label{varsimple}
\end{equation}
This result
has a simple physical interpretation. For power spectra like in
equation (\ref{ps}) the dominant contribution to gravitational deflection 
of light comes from the scales near the
turnover position
$k_0^{-1}$. A photon travelling through a coherent structure
of size $k_0^{-1}$ will be deflected by $\delta \gamma 
\approx 2\phi \approx 6Q_2$,
where the last relation assumes $\phi$ is scale invariant for 
$k<k_0$ and is therefore fixed by the Sachs-Wolfe effect on the Hubble 
distance scale. 
Each region of size $k_0^{-1}$ makes an independent contribution
to the deflection. Since the individual contributions are random,
the photon exhibits a random walk with $\sigma_{\gamma} \approx
N^{1/2} \delta \gamma $, where $N=k_0r_{OS}$. Numerical factors aside
this agrees with equation (\ref{varsimple}).
A reasonable value for the turnover position in the power spectrum is given by
$k_0^{-1}=10$ Mpc. Taking $r_{OS}=1$ 
Gpc as a typical source distance
we find $\sigma_{\theta} \approx 3\times 10^{-4}(k_0 r_{OS}/100)^{1/2}
\approx 1' (k_0 r_{OS}/100)^{1/2}$. 
This is small compared to 1, which justifies
the small deflection angle assumption. It also verifies that the
pathlengths are not  
significantly lengthened by the perturbations. 

However, we conclude that 
the fluctuations in the angular position $\sigma_{\theta}$
are of the order of 
arcminutes (see also Linder 1990), 
which means that the true positions of distant objects in the
universe, such as quasars, differ from the
measured positions on average (rms) by this amount. 
These fluctuations arise already because of the linear structures (voids
and superclusters) and are
present
even when there are no
nonlinear objects (like galaxies and clusters) intersecting the
photon trajectories. 
The fluctuations are 
much larger than a typical image separation in a 
lens system, which is of the order of a few arcseconds. 
The large total deflections are
not directly observable in a single lens system, because only the relative 
positions of images can be measured. 
Figure \ref{fig1}
shows photon propagation in a typical two image GL system.
Despite the fact that the deflection of any single photon ray can be large, 
the lens equation will still give the same solution
as in the unperturbed case, provided that LSS deflects
the two photons approximately 
by the same amount. This will be examined below.
Although the
deflection of a photon ray
relative to the unperturbed direction is not directly observable from the 
positions, one might worry that it could produce significant 
time delay fluctuations. We will address this question in \S 4.

Let us now calculate LSS effects on the relative image positions,
by calculating $\sigma_{\Delta \gamma}$ and $\sigma_{\Delta \theta}$,
the
dispersion in the relative direction and in the relative 
angular position separation between two image rays. 
We will denote the two rays with 
$A$ and $B$, 
separated in direction at the observer's position by an 
angle $\Delta \gamma_0=\gamma_1^A(0)-\gamma_1^B(0)$, 
where the orientation of the coordinate system was chosen  
so that the image separation vector lies
in the direction $\bi{e_1}$ (so that $\gamma_2^A(0)=\gamma_2^B(0)$).
We divide the potential into a stochastic part, 
which describes the LSS, and a non-stochastic 
part describing the primary lensing object. We assume
there are no correlations between these two parts.
Taking the expectation value of equation (\ref{lenseq}) 
the stochastic contributions average to 0 and 
we obtain the usual lens equation 
in the thin lens approximation.

The difference between the two direction vectors caused by LSS
between the lens and the observer is given by 
\begin{equation}
\Delta \bi{ \gamma}(r_{OL})= \Delta \gamma_0 \bi{e_1}+ 
\bi{\alpha_{\perp}}^A(0,r_{OL})-
\bi{\alpha_{\perp}}^B(0,r_{OL}),
\end{equation}
where
$\bi{\alpha_{\perp}}^A(0,r_{OL})$ and
$\bi{\alpha_{\perp}}^B(0,r_{OL})$ 
are the LSS caused
deflections for the rays $A$ and $B$, respectively. 
We  excluded the non-stochastic
deflection from the primary lens itself. 
For a fixed photon separation angle at the observer's position 
$\Delta \gamma_0$, the
rms fluctuation
in the angle between the two rays at the lens position is given by
\begin{eqnarray}
\sigma_{\Delta \gamma}={1 \over \sqrt{2}}\biggl\langle \biggl[
\bi{\alpha_{\perp}}^A(0,r_{OL})-
\bi{\alpha_{\perp}}^B(0,r_{OL})\biggl]^2
\biggl\rangle^{1/2}=\biggl[C(\Delta \gamma_0 )-C(0)\biggl]^{1/2},
 \label{ad}
\end{eqnarray}
where $C(\gamma)$ is given by equation (\ref{limber}) with $q(r)=1$ and
$r_0=r_{OL}$. 
Assuming in addition that the scales contributing to the fluctuations 
are much larger than the typical separation between the two rays (one can
always verify this assumption by evaluating $\sigma_{\Delta \gamma}$
from equation
\ref{limber}), we obtain
\begin{equation}
\sigma_{\Delta \gamma}=2\pi\Delta \gamma_0 
\biggl[{r_{OL}^3 \over 3}
\int_0^{\infty}P_{\phi}(k)k^5
dk\biggl]^{1/2}. 
\label{dtfinal}
\end{equation}
Replacing $\Delta \gamma_0$ with 
$(r_{OL}/r_{LS})\Delta \gamma_0$ and $r_{OL}$ with $r_{LS}$
in equation \ref{dtfinal}
gives the rms fluctuation 
between the two ray directions accumulated between the source and 
the lens. Summing the two contributions in quadrature gives  
the rms fluctuation accumulated between the source and the observer, 
neglecting the correlations between paths on either side of the lens.
Similarly, the dispersion
of the angular position in the lens plane is given by setting 
$q(r)=(1-r/r_{OL})$, which gives for the contribution between the
observer and lens,
\begin{equation}
\sigma_{\Delta \theta}={\sigma_{\Delta \gamma} \over \sqrt{10}}=
2\pi\Delta \gamma_0 \biggl({r_{OL}^3 \over 30}
\int_0^{\infty}P_{\phi}(k)k^5
dk\biggl)^{1/2}.
\end{equation}

For the simple power spectrum of equation (\ref{ps})  
we find
\begin{equation}
\sigma_{\Delta \gamma}=
\sqrt{10}\sigma_{\Delta \theta}
\approx 4 Q_2(k_0r_{OL})^{3/2} \Delta \gamma_0 
\approx 0.025 (k_0r_{OL}/100)^{3/2}
\Delta \gamma_0.
\label{difs}
\end{equation}
Again, there is a simple physical explanation of this result.
Two photons separated by an angle $\Delta \gamma_0$ sample different
potentials, $\delta \phi \approx (\nabla_{\perp} \phi) r\Delta \gamma_0 \approx
k_0\phi r\Delta \gamma_0$, including only the peak power 
contribution around $k_0$. 
The separation $r\gamma_0$ is largest at the lens, but only
falls to one-half for distances half and three-halves as far, so it is a
reasonable approximation to fix $r$ to $r_{OL}$.
A coherent structure of size $k_0^{-1}$ 
leads to an angular difference of $\delta \Delta \gamma \approx 2\delta
\phi$ and there are $N=k_0r_{OL}$ random and independent contributions.
The total angular 
difference is just an incoherent sum of individual contributions,
$\sigma_{\Delta \gamma} \approx N^{1/2}\delta \Delta \gamma$,
which, numerical factors aside, reproduces equation (\ref{difs}).

A more direct way to estimate the amplitude of image position 
fluctuation is to use
the observations of correlated distortions of distant galaxy images. 
This can be described by the ellipticity correlation function 
$C_{pp}(\gamma)$ (Blandford et al. 1991), 
which describes the correlations in the ellipticities of galaxy 
images as a function of angular separation $\gamma$. 
The 
ellipticity correlation function at zero lag, $C_{pp}(0)$, can be 
related to the power spectrum using the expression
\begin{equation}
C_{pp}(0)={16\pi^2r_g^3 \over 30}
\int_0^{\infty}P_{\phi}(k)k^5 dk,
\end{equation}
where we assumed for simplicity that all the galaxies lie at the same 
distance $r_g$ (Kaiser 1992; Blandford et al. 1991). 
From this one sees that 
\begin{equation}
{\sigma_{\Delta \theta}\over \Delta
\gamma_0}={\sigma_{\Delta \gamma} \over \sqrt{10} \Delta
\gamma_0}=
{1 \over 2} C_{pp}^{1/2}(0)(r_{OL}/r_g)^{3/2}.
\end{equation}

For most cosmological models and distances, the linear theory prediction 
of equation (\ref{dtfinal}) 
gives $\sigma_{\Delta \gamma}/\Delta \gamma_0$ of the order of a few percent. 
This has to be corrected for the nonlinear effects, which 
are somewhat uncertain.
Theoretical estimates (Kaiser 1992) and N-body 
simulations (Blandford et al. 1991; Miralda-Escud\' e 1991)
suggest that $C_{pp}(0)$ is unlikely 
to exceed $10^{-3}$. This is also supported by the observational data.
Mould et al. (1994) report a detection of a signal with 
a value $C_{pp}(1^\prime <\gamma<5^\prime)=(5.6\pm
0.6) \times 10^{-4}$, which, after seeing correction,
implies average ellipticity within
a few arcminutes radius of about 0.05. 
Assuming this value
we find
$\sigma_{\Delta \gamma}/\Delta
\gamma_0 \approx 0.08(r_{OL}/r_g)^{3/2}$, with
$r_g \approx 0.6H_0^{-1}$. 
However, the authors could not exclude the possibility that 
the observed signal  
is due to systematic effects. 
This is also suggested by Fahlman et al. (1994),
who report a null detection of average ellipticity within a $2.76^\prime$
radius aperture with a sensitivity of about 1.3\%, which after adjustment 
to $r_g$ above implies an upper limit  
$\sigma_{\Delta \gamma}/\Delta
\gamma_0 < 0.03(r_{OL}/r_g)^{3/2}$.
These measurements give average ellipticities in typically arcminute size 
windows and do not probe $C_{pp}(\gamma)$ on scales below $1^\prime$. 
Observationally it is difficult to give reliable estimates on smaller scales
because one has to distinguish between the signal and 
the noise from the intrinsic 
ellipticities of galaxies. A rather weak upper limit on $C_{pp}(0)$
can be obtained simply from the average ellipticity of galaxies, 
which is of the order of 0.4 and is
dominated by intrinsic ellipticities. 
Despite some uncertainty from the model predictions and observations,
it appears unlikely that the relative
fluctuations in the image separation angle exceed 
a level of a few percent.  

The conclusion above justifies the assumption
that the rms fluctuation is small compared to the measured image
separation. The fact that $\sigma_{\Delta \gamma}/\Delta \gamma_0
 \ll 1$ also implies 
that one cannot have multiple images produced by LSS alone. 
Therefore, multiple images can only be formed from nonlinear 
structures, such as galaxies or clusters. This conclusion has 
previously been 
obtained using N-body simulations by
Jaroszynski et al. (1991) and using semi-analytical methods by 
Bartelmann \& Schneider (1991). 
The fluctuations in angular image separation, although small, 
are in most cases larger than the observational errors on the image
positions (typically less than 0.01 arcsecond/arcsecond $\sim 10^{-2}$). 
Therefore, LSS effects are a major source of uncertainty in the 
true image positions. Observational efforts in
trying to determine the image position with precisions below 0.01 arcsecond
are redundant and do not improve the lens reconstruction.
This effect fundamentally limits our ability to reconstruct
the lens potential using the image positions and should be included in the
modelling of lens parameters, once the actual amplitude of fluctuations 
is determined from the ellipticity correlation function measurements. 

We should also justify the assumptions made in our calculations. 
One is that the statistical properties
of the potential
along the perturbed path are well approximated by those along 
the unperturbed path, as
expressed in equation
(\ref{assum}). 
Taylor expansion of the potential gives 
\begin{equation}
\phi(\bi{x})\approx \phi(r\bi{\widehat{r}})+\bi{x_{\perp}}\cdot
\bi{\nabla_{\perp} }\phi(r\bi{
\widehat{r}}).
\end{equation}
Inserting this into the 
left-hand side of equation (\ref{assum}), we find that for Gaussian 
random fields the relative
correction to the right hand side of equation (\ref{assum}) is
given approximately by
$(x_{\perp}k_0 )^2 \approx [3 Q_2 (k_0r_{OL})^{3/2}]^2
\approx 10^{-3} (k_0r_{OL}/100)^3$.
Therefore, the approximation in equation (\ref{assum})
leads to negligible errors.  
Another approximation we used was to 
neglect the correlations between LSS and the primary lens.
This is justified because the two are correlated 
only over a correlation length distance, which is much smaller than the 
typical pathlength.
While there are $N$ uncorrelated regions of size $k_0^{-1}$
along the photon path, only one of those is strongly correlated 
with the primary lens. The contribution from that region can be regarded as 
being part of the primary lens itself. 
The error due to this approximation is therefore of the order of
$N^{-1}=(k_0r_{OS})^{-1} \ll 1$.

\section{Fluctuations in time delay}

\begin{figure}[p]
\vspace*{6 cm}
\caption{Schematic diagram of a typical
lensing case, as discussed in the text.
Solid lines represent true photon trajectories, dashed lines apparent 
trajectories as seen from the observer's position and dotted lines 
the unperturbed trajectories as seen from the lens plane
in the absence of LSS effects.
The apparent
image and lens positions are denoted by $A^\prime$, $B^\prime$ and $L^\prime$, 
respectively, and
can be far from the true positions. }
\label{fig1}
\includegraphics{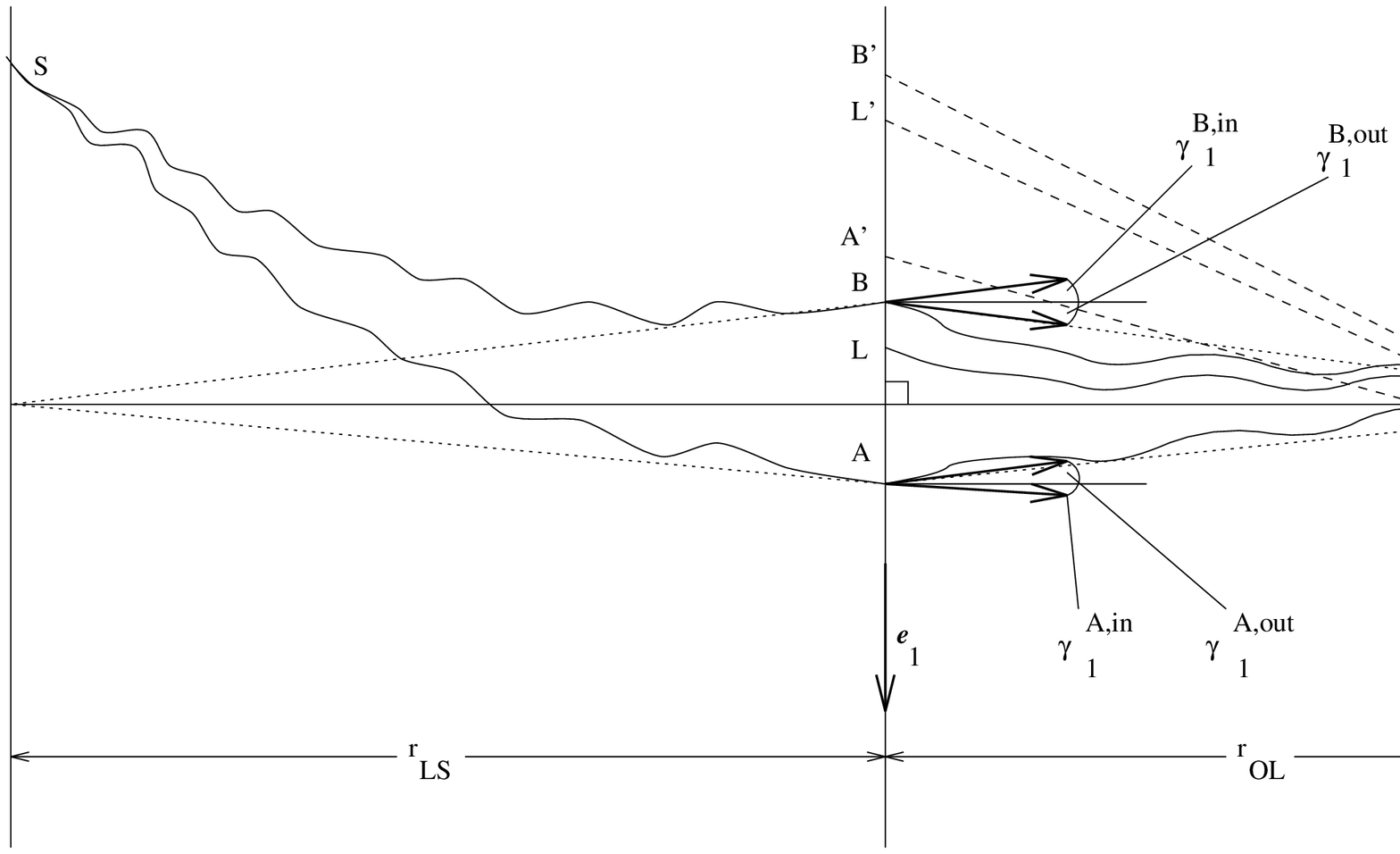}

\end{figure}

In this section we compute
the dispersion in time delay between two images. 
For this purpose 
it is useful to define the time delay relative to the normal of the 
lens plane 
and not relative to the
source-observer line.
We define the lens plane at the lens redshift $z_l$
to be orthogonal to what would be the source-observer line in the 
absence of LSS effects (dotted lines on
Figure \ref{fig1}).  
Relative to the lens plane normal, 
the incoming and outgoing photon direction vectors in the lens plane
are $\bi{\gamma^{A,in}}$, $\bi{\gamma^{A,out}}$ and $\bi{\gamma^{B,in}}$, 
$\bi{\gamma^{B,out}}$ for the images $A$ and $B$, respectively.
The difference between the incoming and outgoing photon direction
gives the deflection angles
in the lens plane, $\bi{\widehat{\alpha}}^A$ and
$\bi{\widehat{\alpha}}^B$. These can 
be obtained by modelling the lens potential using various observational
constraints, such as image magnifications, velocity dispersion of the 
lensing galaxy and/or cluster, positions of other images or 
arcs, etc. 

In the absence of LSS the time delay between two images is given 
from equation \ref{tdgen} by
\begin{eqnarray}
 \Delta t &=&
{1 \over 2}\left\{r_{LS}[(\gamma_1^{A,in})^2-(\gamma_1^{B,in})^2]
+r_{OL}[(\gamma_1^{A,out})^2-(\gamma_1^{B,out})^2]\right\} -
2(\psi^A-\psi^B)
\nonumber
\\&=&
{r_{OL}r_{OS} \over 2r_{LS}}[(\gamma_1^{A,in})^2-
((\gamma_1^{B,in})^2]-
2(\psi^A-\psi^B).
\label{tdtl}
\end{eqnarray}
Here $\psi^A$, $\psi^B$
are the integrals of the primary lens potential for the two rays, 
\begin{equation}
\psi=\int_{r_{OL}-\epsilon}^{r_{OL}+\epsilon}\phi(r)dr
\end{equation}
with $\epsilon/r_{OL} \ll 1$.
Equation \ref{tdtl}
is the usual time delay expression in the thin lens approximation
(e.g. Blandford \& Narayan 1986; Blandford \& Kochanek 1987; 
Schneider, Ehlers \& Falco 1992). 
Note that the difference between the two outgoing photon directions 
gives the observed image splitting.

Adding LSS moves both the source and the observer, for a fixed lens plane 
(Figure \ref{fig1}).
The time delay between the two rays is now given by 
\begin{eqnarray}
\Delta t&=&{1 \over 2}
\int_{r_{OL}}^{r_{OS}}
\left\{[\bi{\gamma}^{A,in}-\bi{\alpha}^A_{\perp}(r_{OL},r)]^2
-[\bi{\gamma}^{B,in}-\bi{\alpha}^B_{\perp}(r_{OL},r)]^2
\right\}dr 
\nonumber  \\
&+&{1 \over 2}
\int_0^{r_{OL}}\left\{[\bi{\gamma}^{A,out}+\bi{\alpha}^A_{\perp}(r,r_{OL})]^2
-[\bi{\gamma}^{B,out}+\bi{\alpha}^B_{\perp}(r,r_{OL})]^2
\right\}dr \nonumber \\
&-&2\left\{ 
\int_{r_{OL}}^{r_{OS}}[\phi^A(r)-\phi^B(r)]dr+
\psi^A-\psi^B+
\int_0^{r_{OL}}[\phi^A(r)-\phi^B(r)]dr\right\}.
\label{tdl} 
\end{eqnarray}
This equation is similar to equation (\ref{tdgen}), except that here the 
geometrical contribution is measured relative to the normal of the lens plane.
The first two lines in equation (\ref{tdl}) give the 
geometrical time delay between 
the lens and the source and between the lens and the observer, 
respectively. In the third line we have written the
gravitational time delay contribution coming from the potential  
between 
the lens and the source, from the primary lens potential and from the 
potential between the lens and the
observer, respectively.
The LSS ($\bi{\alpha}_{\perp}$, $\phi$) and primary lens 
($\bi{\gamma}^{in}$, $\bi{\gamma}^{out}$, $\psi$)
contributions are thus explicitly separated.

Let us calculate the time delay contribution 
between the ray $A$ and the fiducial ray
accumulated 
between the lens and the 
observer. 
The fiducial ray is defined to start perpendicular
to the lens plane and end at the observer's positions. 
The total time 
delay is obtained by adding a similar contribution from 
the lens to the source and subtracting the same terms for the 
ray $B$.
The fiducial ray direction is given by 
\begin{equation}
\bi{\gamma^f}(r)=\bi{\alpha_{\perp}}
(r,r_{OL})=\int_r^{r_{OL}} \bi{g_{\perp}}(r)dr,
\end{equation}
where $\bi{g_{\perp}}(r)$ is computed along the fiducial ray.
The direction of the ray A is 
\begin{equation}
\bi{\gamma}^A (r)= 
\bi{\gamma^f}(r)+\bi{\gamma}^{A,out}+ \Delta\bi{\alpha}^A_{\perp} 
(r,r_{OL}).
\end{equation}
Here $\Delta\bi{\alpha}^A_{\perp}
(r,r_{OL})$ is the difference between the LSS induced 
ray deflections at $r$ and can
be calculated using the Taylor expansion of $\bi{g_{\perp}}$ around the 
fiducial ray, as we did in \S 3.
The image ray position relative to the fiducial ray is
\begin{equation}
\bi{x}^A_{\perp}(r)=\bi{x}^A_{\perp}(r_{OL})+\bi{\gamma}^{out,A}(r_{OL}-r)+
\int_r^{r_{OL}}
\Delta \bi{\alpha}^A_{\perp} (r',r_{OL})dr'.
\end{equation}
The initial lens plane position of the image ray relative to the 
fiducial ray, $\bi{x}^A_{\perp}
(r_{OL})$,
is not a free parameter, since it
has to satisfy the constraint 
\begin{equation}
\bi{x}^A_{\perp}(0)=0.
\end{equation}
From this we obtain 
\begin{equation}
\bi{x}^A_{\perp}(r)=-\bi{\gamma}^{out,A}r-\int_0^r\Delta 
\bi{\alpha}^A_{\perp} (r',r_{OL})dr'.
\end{equation}

The gravitational time delay contribution is obtained from the Taylor 
expansion of the potential around the fiducial ray, 
\begin{equation}
\Delta t_{grav}=\int_0^{r_{OL}}\left\{\bi{g_{\perp}}(r)\cdot \bi{x}^A_{\perp}(r)
+O[(\bi{x}^A_{\perp}(r)^2]\right\}dr.
\label{dtgra}
\end{equation}
The geometrical time delay contribution is given by 
\begin{eqnarray}
\Delta  t_{geom}&=&{1 \over 2} \int_0^{r_{OL}}[\bi{\gamma}^A(r)^2-
\bi{\gamma^f}(r)^2]dr 
\nonumber\\
&=&{\bi{(\gamma}^{out,A})^2r_{OL} \over 2} + 
\int_0^{r_{OL}}\biggl[
\bi{\gamma^f}(r)\cdot \bi{\gamma}^{out,A}
 + \bi{\gamma^f}(r) \cdot \Delta \bi{\alpha}^A_{\perp} 
(r,r_{OL}) 
\nonumber\\
&+& \bi{\gamma}^{out,A} \cdot 
\Delta \bi{\alpha}^A_{\perp} (r,r_{OL}) + {\Delta \bi{\alpha}^A_{\perp} 
(r,r_{OL})^2 \over 2}\biggl]dr.
\end{eqnarray}
Integrating by parts
the terms involving $\bi{\gamma^f}(r)$ we find 
\begin{eqnarray}
\int_0^{r_{OL}}\bi{\gamma^f}(r) \cdot [\bi{\gamma}^{out,A}+\Delta
\bi{\alpha}^A_{\perp} (r,r_{OL})]dr= \nonumber \\
\int_0^{r_{OL}}\left\{\bi{g_{\perp}}(r) \cdot \int_0^r[\bi{\gamma}^{out,A}
+\Delta
\bi{\alpha}^A_{\perp} (r',r_{OL})]dr'\right\}dr.
\end{eqnarray}
This is exactly 
cancelled by the first order term in $\Delta t_{grav}$ (equation \ref{dtgra}). 
Therefore, $\bi{\gamma^f}$ completely drops out of the time delay 
expression and 
using time delay measurements we cannot infer any information on the 
absolute deflection angle. This is quite remarkable, given that 
separately the geometrical and gravitational LSS induced fluctuations are
approximately $ 15 \ \mbox{yr}\ (k_0r_{OS}/100)^{1/2}
(r_{OS}/1\mbox{Gpc})(\Delta \gamma_0/1'')$, 
much larger than the expected
time delay from the primary lens itself, of the order of
$0.1\ \mbox{yr}\ (r_{OS}/1\mbox{Gpc})(\Delta \gamma_0/1'')^2$.

Adding the geometrical and gravitational time delay contributions
we finally obtain
\begin{equation}
\Delta t^A={(\bi{\gamma}^{out,A})^2r_{OL} \over 2}+\int_0^{r_{OL}}
[\bi{\gamma}^{out,A}\cdot 
\Delta \bi{\alpha}^A_{\perp} (r) + {1\over 2}\Delta \bi{\alpha}^A_{\perp}
(r)^2+O((\bi{x}^A_{\perp}(r)^2)]dr.
\label{tdp}
\end{equation}
The first term in equation (\ref{tdp}) 
gives the largest contribution and is the term that one would also have in the
absence of LSS (compare with equation 
\ref{tdtl}). The second term is smaller than the 
first term approximately by 
$\Delta \alpha_{\perp}^A(0,r_{OL})
/\bi{\gamma}^{out} \sim \sigma_{\Delta \gamma}/
\Delta \gamma_0 $. The last two terms in equation
(\ref{tdp}) are further 
suppressed by $\sigma_{\Delta \gamma}/
\Delta \gamma_0$
relative to the second term and can be neglected.

What, then, is 
the LSS induced fluctuation that causes the reconstructed time delay 
to differ from 
the true time delay? The observer measures the image 
separation angle that 
is almost, but not exactly, given by $\bi{\gamma}^{A,out}-\bi{\gamma}^{B,out}$,
so that the reconstructed time delay 
differs somewhat from equation (\ref{tdtl}). 
It is not possible to give an exact 
prediction of the reconstructed time delay without specifying the detailed
lens model and taking into account all of the observational constraints. 
It is clear, however, that the fluctuation in the reconstructed time delay 
is due only to the fluctuation in the relative angle separation, part of
which is described by the second term in equation (\ref{tdp}).
Given that 
$\sigma_{\Delta \gamma}/ \Delta \gamma_0 \ll 1$, the relative
effects on the
time delay will also be of that order. 
We conclude
that the time delay fluctuation
induced by LSS is of the order of $\sigma_{\Delta \gamma}/ \Delta \gamma_0$,
which is of the order of a few percent for sources and
lenses at cosmological distances. 

\section{Conclusions}

We investigated the LSS effects on measurable properties of gravitational
lens systems, in particular on the image positions and time delays.
The method we use is based on a geodesic equation in a weakly perturbed
flat
Robertson-Walker metric and is valid for a general matter distribution 
between the source and the observer. 
The advantage of this approach compared to previous work on this 
subject is that it only assumes the knowledge of evolution of
density power spectrum, which can easily be related to other 
measurements of LSS to obtain quantitative predictions. 
The same approach can also be used 
to calculate light propagation in 
non-flat universes and the conclusions in this paper do not 
significantly depend on the assumed value of $\Omega$. 
We find that the
rms fluctuation in the relative positions of images is $\sigma_{\Delta \gamma}/
\Delta \gamma_0 \sim 0.025(k_0 r_{OL}/100)^{3/2}$, for a LSS density 
power spectrum peaking at wavelength $k_0$. For most realistic models of 
LSS this is much
smaller than unity and so one
does not expect multiple images generated 
from LSS. 
Nevertheless, these fluctuations are likely to be 
larger than the observational errors
and should be included in the modelling of lens
parameters.  
Similarly, rms fluctuation in time delay due to LSS 
are caused only by the uncertainties in the relative 
image positions and 
are also approximately given by $
0.025(k_0 r_0/100)^{3/2}$. Therefore, LSS does not significantly
affect the time delays.
While the same method can be used to predict fluctuations in the relative
image magnification and orientation, a simple estimate shows that the 
effect on these observables 
is negligible. The rms fluctuation in relative magnification between two 
images $\Delta M/M$ is
given approximately by
$\Delta M/M \sim (\nabla_{\perp} M) r_{OS}\Delta \gamma_0/M \sim 
(k_0r_{OS}\Delta \gamma_0)C_{pp}(0)^{1/2} \sim 10^{-5}$,  
well below the measurement errors. 

The conclusion regarding LSS effects on time delay measurements 
disagrees with the conclusions reached by 
Alcock \& Anderson (1985, 1986), Watanabe et al. (1992) and 
Sasaki (1993). Sasaki (1993) suggests that the coupling between 
the absolute and relative deflection, which in our language is
proportional to $\sigma_{\gamma}\sigma_{\Delta \gamma}$, generates
large time delay fluctuations. As we have shown in \S 4, this term
actually vanishes once both the gravitational and geometrical time delay
contributions are included. 
Alcock \& Anderson (1985, 1986),
Watanabe et al. (1992) and Sasaki (1993) have argued that although
the universe is homogeneous on large 
scales, individual photons may travel through a region
where the density differs from the 
average density in the universe. 
If one assumes that the density in the beam is some constant fraction of the 
average density, then
one can use the Dyer-Roeder angular 
distance redshift relation (see Sasaki 1993 for a comprehensive discussion)
to deduce the value of the Hubble constant.
The result depends on the unknown density in the beam and one may
conclude that this significantly affects the Hubble constant determination
from the time delay measurements.
For example, Alcock \& Anderson (1985, 1986) argue that because of LSS
correlations there
may be a significant overdensity in the particular 
direction of the lens.  
This conclusion 
neglects the fact that LSS and the primary lens are correlated
over only a correlation length distance, which is typically 
much smaller than the photon travel distance.
Similarly, Watanabe et al. (1992) and Sasaki (1993)
argue that we might live in a universe where
part or all of the mass is concentrated in small clumps
away from the photon beam, which are
unable to significantly affect the light propagation.
Both descriptions of LSS are oversimplified in that they
do not account for the stochastic nature of LSS. 
Dynamical measurements over the last decade show large amounts
of dark matter present on all scales and
so one cannot assume that photons are travelling through a
uniformly filled or empty beam when away from the primary lens. 
Moreover, the average 
redshift-angular distance relation should not 
differ from the homogeneous case, once we average over all lines of sight 
(including those through the clumps of matter). 
If redshift-angular distance relation for
most lines of sight is to differ significantly from 
that in the homogeneous universe, 
then this requires
large rms fluctuation around the mean.
This would imply large 
$\sigma_\gamma/\gamma_0$ but,
as we argued in \S 3, no such large fluctuations are either expected or
observed.
Therefore, the 
use of angular distances as given by the homogeneous universe model
appears to be adequate. 
Another reason for the disagreement with
Watanabe et al. (1992) is that these authors assumed a large contribution
to the fluctuations from small (galactic) scales.
As we argued, the data at present show little support for such a large
contribution on those scales, although for a more quantitative prediction
better measurements of power spectrum on small (subarcminute) scales
would be needed.

If LSS does not induce significant time delay fluctuations,
then this would remove one of the major objections against using time 
delay measurements to determine $H_0$. 
Significant problems 
related to the robustness of the lens reconstructions still
remain and are preventing the method at present
from giving a reliable estimate of $H_0$ (see e.g. Bernstein, 
Tyson \& Kochanek 1993 for a discussion of lens reconstruction in 0957+561). 
Moreover, the above analysis does not exclude the possibility that 
a homogeneous sheet of matter is present in the lens plane, 
because in our treatment this sheet of matter is part of the primary lens.
Uniform matter distribution, which is likely to be overdense close to the
primary lens, cannot be determined from the image positions,
but it does affect the length scale and 
makes the deduced value of Hubble constant larger than the true value 
(Borgeest \& Refsdal 1984; Falco, Gorenstein \& Shapiro 1991). 
Although severe,
these problems are not unsolvable and GL time delay method 
remains one of the few
methods that can provide information on the global distance scale and 
geometry of the universe.

\acknowledgements
I would like to thank Ed Bertschinger for 
a careful reading of the manuscript and useful discussions.
I would also like to thank Richard Gott, Jim Gunn, Lam Hui, Nick Kaiser, 
Jordi Miralda-Escud\' e, Ramesh Narayan, 
Jerry Ostriker, Misao Sasaki, Ed 
Turner, Kazuya Watanabe and
the participants of the 
seminar on gravitational lenses at MIT for useful discussions.
This work was supported by grants NSF AST90-01762 and NASA NAGW-2807.

\end{document}